**OVIDIU ANDREI SCHIPOR, FELICIA GÎZĂ BELCIUG, ŞTEFAN-GHEORGHE PENTIUC, CRISTIAN EDUARD BELCIUG, MARIAN NESTOR**

# SOFTWARE PACKAGE WITH EXERCISES FOR THERAPY OF CHILDREN WITH DYSLALIA

*"Ştefan cel Mare" University of Suceava,*
*University street 13, 720229, Suceava, Romania,*
*tel.: +40.230.524.801, E-mail: schipor@yahoo.com*

**Abstract.** In this paper we present a consistent set of exercises for children with dyslalia (dyslalia is a speech disorder that affect pronunciation of one ore many sounds). The achievement has gone from "Therapeutic Guide" made available by the team of researchers led by Professor Mrs. Iolanda TOBOLCEA from the "Alexandru Ioan Cuza" University of Iasi. The specifications of the "Therapeutic Guide" have been fully complied with, such exercises being adapted both age and level of children.
To achieve these exercises were recorded, processed and used more than 10000 voice production. They also have been made over the 2000 corresponding recorded nouns. These exercises are being tested by Interschool Regional Logopaedic Center of Suceava - Romania.

## INTRODUCTION

In this article we refer to LOGOMON system developed in TERAPERS project by the authors. The full system is used for personalized therapy of dyslalia affecting pre scholars (children with age between 4 and 7). Dyslalia is a speech disorder that affect pronunciation of one ore many sounds. According to the statistics, about 10% of pre scholars are affected by this type of speech impairment [1].

To be assisted instruction children dyslalia both in the speech therapy laboratory and in the family, prepared a set of specific exercises. The achievement has gone from "Therapeutic Guide" made available by the team of researchers led by Professor Mrs. Iolanda TOBOLCEA from the "Alexandru Ioan Cuza" University of Iasi. The specifications of the "Therapeutic Guide" have been fully complied with, such exercises being adapted both age and level of children.

For effective implementation of the exercises were necessary [2][3]:
- description of specific steps for each exercise;
- indicating speech productions that can be content exercises (syllabi, divided words, words related, appropriate words, structures through progressive addition, sentences, etc);
- speech productions recording by specialized personnel (speech therapists from Interschool Regional Logopaedic Center of Suceava);
- collected from various sources visual material support (for example exercisesthat implies nouns);
- designing and implementing a database allowing storing:
  - multimedia elements necessary for exercises;
  - exercises (sets of multimedia elements);
  - homework (sets of exercises);
- development of software interfaces for the introduction of multimedia elements (vocal production and images) in the database;
- carrying out a way to create exercises and homework;
- the possibility of transferring these exercises and homework to be run on the child PC or PDA.

The first two issues were offered by "Therapeutic Guide" prepared by UAIC Iasi. The records were made necessary conditions studio audio by speech therapists respecting the particulars required by dyslalia (emphasis in the pronunciation of sounds that are willing to be corrected, keeping a low tempo in order to increase of comprehensibility). All aspects of the software involved were developed at the "Stefan cel Mare" University of Suceava and tested by CLIJ Suceava and UAIC Iasi.

Software modules that allow developing and running exercises have been integrated into the LOGOMON system [4]. In this way, speech therapist has the possibility of a unified control over all aspects that are involved in child therapy.

It should be noted that the results of child issues are not used only by speech therapist. They are stored in the database and provide the opportunity to preserve a history of child development [5]. Starting from these historic conclusions can be drawn with therapeutic implications best route to be followed in future. In addition to the possibility of establishing themes by speech therapist, there are variant in which the theme to be given to solving the child is automatically generated by the system, based on existing data. These data are contained on one side in the individual report and on the other hand, in the form of child recordings scores.

Realization and management exercises and themes are consistent manner visually intuitive. Speech therapists that involved in research have learned to use interface in a short time. On the other hand, children were excited to participate in a therapy based multimedia techniques [2].

The difference between multimedia therapies and classical therapies will be established with the help of an experiment (pre-test/post-test of control and experimental groups) [6].

### INTERFACES FOR INTRODUCING MULTIMEDIA ELEMENTS

To introduce multimedia elements (sounds and images), must access menu Administration – Exercise's elements, according to the Figure 1.

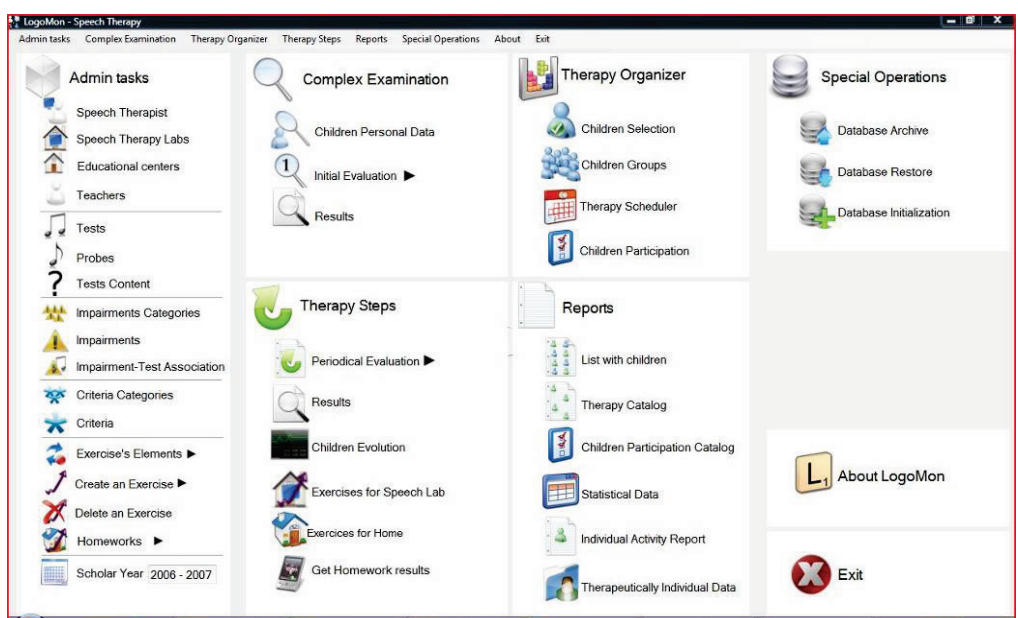

Fig. 1. LOGOMON menus

Creating a word requires two stages, because it was felt that there should be the possibility of editing (change) the parameters of a word. The first window provides a list of all the words in the database, together with speech therapists. When selecting a noun which has an associated image, this image is automatically displayed. To facilitate finding a word in the list, there is a specific search field (*Cauta [Search]*). Add button (*Adauga [Add]*) causes the opening of another window (second window) that allows the introduction of a new word. Edit button (*Editeaza [Edit]*) allows modifying the parameters of a word with an interface similar with that of the *Add* button. It is also possible to delete a word from the database.

The main parameter for a word is corresponding audio file (*Fisier audio [Audio file]*). When word is difficult to pronounce, there is the possibility to add an additional audio file, with a slow tempo pronunciation (*Fisier audio despartit in silabe* [Syllabi divided audio file]). Also, for the words that support graphical representations (certain nouns), speech therapist may indicate an image file (*Fisierul imagine [Image file]*).

The user must specify the textual representation of word, on the basis of which can be selected later (at creating exercises) words with certain sounds. The text of the first syllabi (*Text prima silaba*) is necessary for specific exercises.

The words part of speech (e.g. noun) is very important in order to play exercises. Also, there are other parameters required (male or female, articulated or not) that allows automatic formulation requirements in the exercises. Save button (*Salveaza*) and closing button (*Inchidere*) have well known meaning.

The main types of vocal productions used in the exercises are:

- syllabi (e.g. sa-, -as, -asa-, -sas-);
- words (e.g. soare [sunny], casă [hose], vas [boat]);
- appropriate words (e.g. muscă-muşcă [fly-bite])
- onomatopoeia (e.g. the snake: sss – sss - sss)
- mono-syllabi strings (e.g. sa, se, si, so, su, să, sî)
- addition verbal structures (e.g. s, su, sun, sune, sunet, sunete)
- sentences (e.g. Sandu este sănătos. [Sandu is healthy]).

In order to exemplify these facilities, figure 2 show two forms used for words creation.

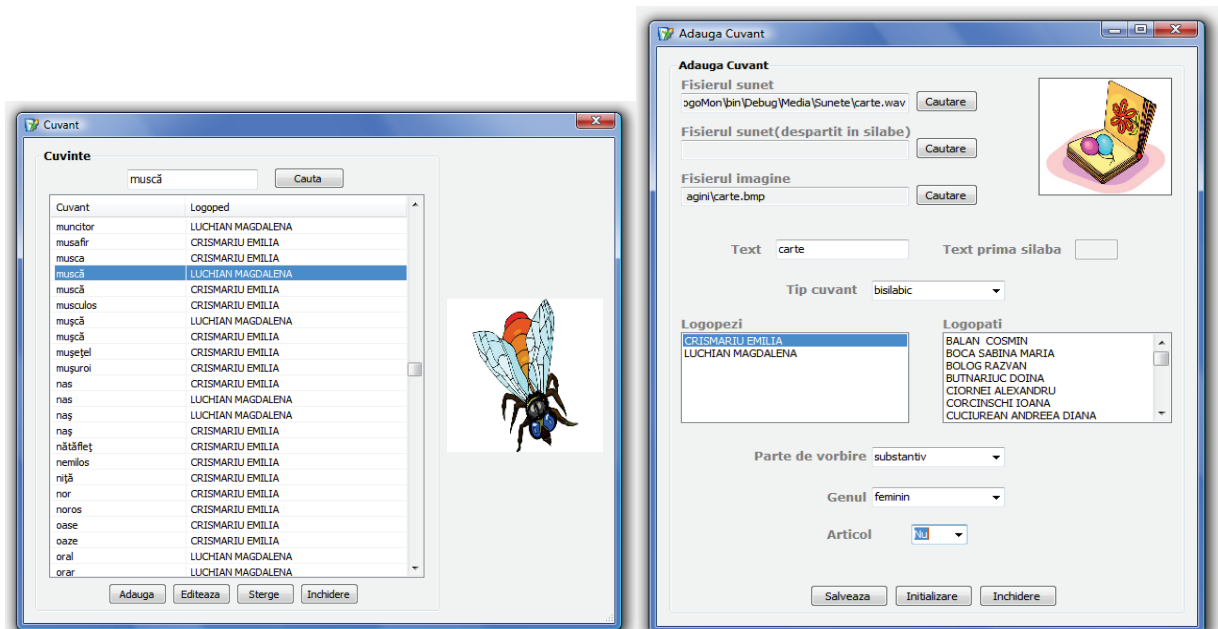

Fig. 2. Forms for creation of words

## INTERFACES FOR EXERCISES CREATION

After multimedia elements (sound and images) are introduced, speech therapist has the ability to create specific exercises for dyslalia by choosing *Create an exercises* option from the *Admin Tasks* menu (Figure 1).

The types of exercises presented in this project are:
- Sound recognition in words - this type of exercise allows the identification of the level in which the child hears the sound concerned.
- Word Identification from pairs of similar words
- Pronunciation of a sound in words and syllables
- Onomatopoeia - After playing indication (for example: *„The train makes şşş-şşş, şşş-şşş. Make you the same."*), the child must imitate the heard pronunciation.
- Verbal structures through progressive addition
- Sounds with similar joints
- Transforming the words by replacement some sounds or syllables

Follow, we present an example of exercise creation ("Intruder recognition"). The user selects the sound for which the exercise will be created (Sunet Exercitiu [Sound of the Exercise]). The degree of difficulty (1-5) will allow further the speech therapist to choice exercises depending on the child. In addition there can be added a literally description of the exercise. Because not all children affected by the system know how to read, the indications are played also in audio format. For example, the selected audio file in the next image contains the words „Click on the image when the sound of R contains the word ".

Subsequently, the words must be selected from the median list, indicating for each word the time for waiting correct action (1-10 seconds) and if it contains the sound concerned or not. Save button (*Salveaza*) causes transposition of the exercise in database. Buttons *Initializare [Init]* and *Inchidere [Close]* brings the window in its original state or close it.

The "recognition intruder" exercise is the most straightforward in which only one word (the intruder) does not contain the concerned sound. This type of exercise has a higher degree of difficulty than others because the discrimination has to be done between two sounds very close (for example: R and L) in similar words (for example: Rac – Lac [Cancer-Lake]). Exercise is implemented in two versions: with images (for similar words which can be graphically represented) and pennants.

The image version: Initial indication is given (for example: Choose the indicated word from the following pairs) both as text and sound. Subsequently are presented successively more pairs of similar words (visual and audio) and the child has to choose a particular word from the pair. He makes the choice by pointing with the left mouse click the correct word specified by computer.

Version with pennants: Initial indication is given (for example: Click on the first flag when it hears the sound of R in the first word and click on the second flag when it hears the sound of R in the second word) both as text and sound.

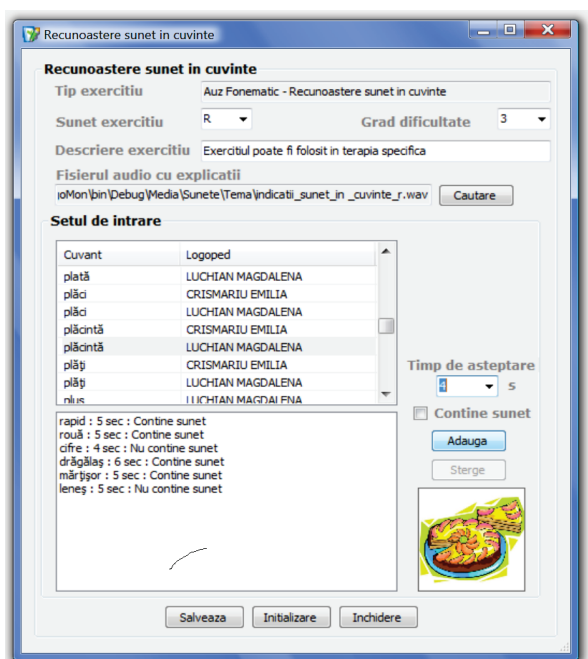

Fig. 3. Recognition of intruder exercise

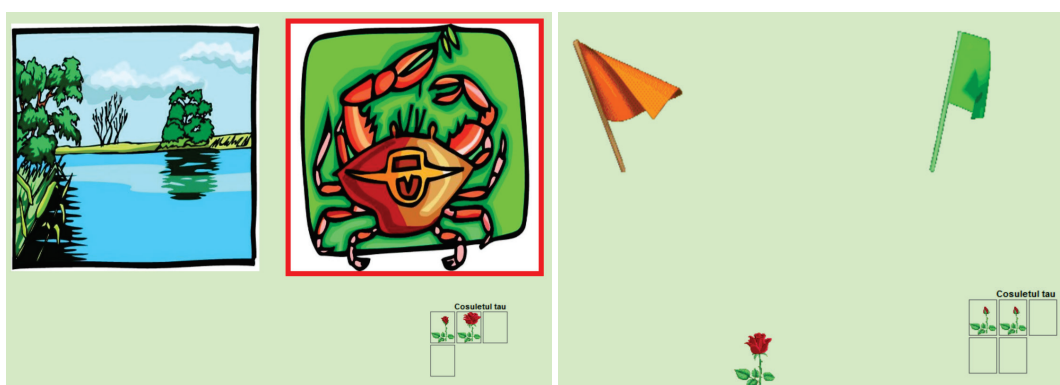

Fig. 4. Running the "recognition intruder" exercise

Subsequently are presented successively more pairs of similar words (audio only) and child must point to the first or second pennant.

Each pair has an associated time (in seconds), which represents the interval left the child to take the correct decision. The passage of time is represented by a progress bar. Each correct choice is rewarded by adding small flowers in child pannier. After passing the list of pairs, are displayed again the pairs for which the child has gave wrong answers.To introduce multimedia elements (sounds and images), must access menu Administration – Exercise's elements, according to the figure 1.

**ACKNOWLEDGEMENTS**

We must specify that these researches are part of TERAPERS project financed by the National Agency

**SCHIPOR OVIDIU-ANDREI** - professor in Electrical Engineering and Computer Science, "Stefan cel Mare" University of Suceava, Suceava, Romania.

**STEFAN-GHEORGHE PENTIUC** - professor on the Faculty of Electrical Engineering and Computer Science University "Stefan cel Mare" of Suceava, Suceava, Romania, tel. +40.230.524.801, http://www.eed.usv.ro/~pentiuc, E-mail: *pentiuc@usv.ro.*

**GIZA-BELCIUG FELICIA** - asistent in *Electrical Engineering and Computer Science*, "Stefan cel Mare" University of Suceava, Suceava, Romania.